\documentclass[10pt,twocolumn]{article} 
\usepackage{latex8}
\usepackage[latin1]{inputenc} \usepackage[T1]{fontenc}
\usepackage{textcomp} \usepackage[british,german]{babel}
\usepackage{mathptmx,courier,xspace}
\newif\ifpdf
\ifx\pdfoutput\undefined\pdffalse     
\else\pdfoutput=1\pdftrue\fi          
\ifpdf                                
  \usepackage[pdftex]{graphicx}       
  \usepackage[pagebackref]{hyperref}  
  \hypersetup{pdfauthor= {Andreas U. Schmidt}}
\else
  \usepackage[dvips]{graphicx}        
  \usepackage[dvips,pagebackref,breaklinks=true]{hyperref}
\fi
\providecommand{\href}[2]{#2}

\providecommand{\hypersetup}[1]{}
\providecommand{\url}[1]{#1}
\newcommand{\CA}{\textsf{A}\xspace}
\newcommand{\CB}{\textsf{B}\xspace}
\newcommand{\VS}{\textsf{VSec}\xspace}
\newcommand{\Arc}{\textsf{Arc}\xspace}
\newcommand{\TTA}{\textsf{T1}\xspace}
\newcommand{\TTB}{\textsf{T2}\xspace}
\begin{document}
\selectlanguage{british}
%
%
\title{A secure archive for Voice-over-IP conversations
\thanks{\copyright ACM, (2006). This is the author's version of the work.
It is posted here by permission of ACM for your personal use. 
Not for redistribution.
The definitive version was published in Proceedings of VSW06, June, 2006, Berlin, Germany.
\url{http://doi.acm.org/10.1145/xxxxxx.xxxxxx}}
}
\author{Christian Hett, Nicolai Kuntze, Andreas U.\ Schmidt\\
Fraunhofer--Institute for Secure Information Technology SIT\\
Rheinstraße 75, 64295 Darmstadt, Germany\\
\texttt{andreas.u.schmidt@sit.fraunhofer.de}}
\maketitle
\thispagestyle{empty}
\begin{abstract}
An efficient archive securing the integrity of 
VoIP-based two-party conversations is presented.
The solution is based on chains of hashes and 
continuously chained electronic signatures.
Security is concentrated in a single, efficient 
component, allowing for a detailed analysis.
\end{abstract}
%
%
\Section{Introduction}\label{sec:introduction}
To archive voice communication in a business context, wherever it is legally possible,
is attractive from a general viewpoint.
The archiving of voice conversations provides inherent evidentiary value 
due to the possibility of forensic evaluation and analysis of the contained biometric data.
Methods for the latter are advanced~\cite{Forensic_Voice_Book}, obtaining to recorded voice 
communication a rather high probative force, e.g., in a court of law.
In comparison to other digital media, e.g., text documents, specific features of voice 
communication can be viewed as contributing to security.
The medium of communication here consists in a linearly time-based full duplex channel
enabling inter- and trans-activity~\cite{Goodwin1981}.
In particular, interactivity enables the partners to make further enquiries in case
of insufficient understanding. That is, communication faults on technical as well as
language levels can be remedied, or at least mitigated, within the ongoing communication.
Furthermore, digital voice communication offers a rather high reliability and quality
of service, leading generally to a higher understandability of VoIP communication
in comparison with its analog predecessors~\cite{Athina2002,hoene2004spects}.
The mentioned properties mitigate to some extent the presentation problem 
to which digital documents are usually prone~\cite{AUS00A}, e.g., misinterpretations due to
misrepresentation, lack of uniqueness of presentation, and inadvertent or malicious hiding
of content.

It is also worth to recall that security of transactions must be assessed in context and can
in general not be reduced to information security.
The communication medium voice contributes to
non-repudiation by offering an independent means of speaker identification.
The voice archiving system presented below is, with respect to non-repudiation,
the analogue of an archive of digital documents which are not signed but only 
time-stamped at entry.
Such a kind of document repository is most commonly used for electronic mail rooms
in the domain of E-government~\cite{Bundesamt2003} 
and records management~\cite{Sprehe2005}.
But due to the intrinsic properties of voice,
the probative value of securely archived conversations can 
reasonably be assumed to be much higher than in this analogy.

The state-of-the art of digital voice recording as used widely for instance 
in financial institutions, is marked by solutions that directly capture VoIP
streams and use security only at the transport layer (SSL)~\cite{ASC,Barker1999}.
The neglect of security of the stored conversations
is perhaps an outcome of the mentioned advantageous features of voice.
Nevertheless, we argue that the probative force of digitally stored voice 
requires in  particular the proper consideration of the integrity
of the voice communication during and after storage, like for any other digital 
data. 

Related work on securing the integrity of streamed data by signatures is scarce.
The authors of~\cite{Perrig2000} describe a method for stream signatures for broadcast 
media. In~\cite{MK1,MK2} a method to transport authentication information employing
watermarks and steganography is presented. Digital signatures are not explicitly used
and achievable data rates seem low.

This motivates our present approach to devise a \textit{secure} archive 
for VoIP-based communication. Section~\ref{sec:requ-secure-arch} explains the
specific security requirements for such an archive. Section~\ref{sec:design-impl}
describes its design and implementation proper, from  base concept and architecture
to implementation. Section~\ref{sec:secur-cons}
examines the most important possible attacks and how they can be dealt with,
thus providing a security analysis of the archive concept.
Section~\ref{sec:conc} contains conclusions and an outlook to future work.
\Section{Requirements for a secure VoIP archive}
\label{sec:requ-secure-arch}
The secure archiving voice based communication must take four main areas of requirements
into account, namely security, efficiency, scalability, and long-term aspects.

In the area of security the main goal is to establish the limited
non-repudiation of the archived communication described in the introduction. 
This comprises
\begin{description}
\item[cohesion]Each archived conversation has to provide a proof of cohesion. 
That is, the ordering, temporal sequencing,  
and completeness of the stored communication packets must be verifiable to enable
a reconstruction of the conversation.
\item[integrity] has to be assured to maintain 
that a communication was not changed at any point during or after archiving.
\item[creation time]Each conversation has to be reliably associated with a certain time,
which must be as close as possible to the conversation's start and the initiation of the
archiving. 
\end{description}
While a secure assignment of a creation time 
is a general requirement for most digital archives, it serves, for a VoIP archive, also
as a base to establish cohesion by providing a temporal context.
Integrity of voice communication refers not only to raw data, but also to cohesion.
Therefore, to achieve the desired combination of cohesion and integrity, it is insufficient
to simply store raw VoIP streams in the archive, since those are amenable to forgery by cutting.

Thus it is highly desirable, both from a security as well as an efficiency
viewpoint, to secure and archive a VoIP conversation as ``close'' as possible
to its transmission, and conceptually close to the actual VoIP stream.
To provide a solution for a secure VoIP archive in this extended sense 
is our main contribution.

The second area regards vital efficiency aspects of a real world
implementation. 
Simplicity of the implementation should minimise 
the effect on existing systems and infrastructures, e.g.,
client-side requirements for the archiving process
should be completely avoided. 
A tight integration is required to enable the utilisation of existing 
infrastructures without or with only minor changes. 
An efficient use of memory and computational resources can be achieved 
by basic conceptual design decisions. 
One major requirement in this context is that the 
data are streamed direct to the archive without necessitating a buffering
of a whole conversation (e.g. to sign it after termination). 
This entails that all security-related operation are
performed on the data stream on-the-fly and optimally 
concentrated in a single module with minimum storage of its own.
Alternatives, like converting a recorded conversation, 
e.g.,to a Blob in the same or another audio format
and signing it after the call is completed 
is conceptually not different from electronically signing an
arbitrary digital document.
Such an approach will a) possibly (depending on the target format) 
loose the contextual information about temporal order and direction of the
communication, and b) introduce another component (for audio conversion) potentially
subject to forgery attacks. Apart from that, it would be 
inefficient since most VoIP codecs already provide good compression and thus
conversion to, e.g., MP3, will be costly in terms of memory and computational power.

The last area is the scalability of the concept. 
Considering the amount of voice
communications in a company or call-centre, 
it is obvious that archiving solutions have to cope with a broad spectrum 
of workloads to be handled. The concept should therefore for instance
enable the usage of external archiving infrastructures.
Scalability is another reason to prefer a streaming security solution
over an intermediate storage and securing conversations afterwards.
In particular the memory requirements in the latter case 
pose, e.g., a hard upper threshold for the number of concurrent calls, respectively 
duration of stored conversations in the temporary memory.

Some long-term aspects of information security  have to be considered as well 
with respect to the archiving and verification processes. 
For instance, the concept has to assure that it is possible at later times to
apply certain transformations on the data as they can be required if, e.g., certificates
are withdrawn, the security of cryptographic algorithms is no longer assured,
 or it is needed to transform the voice data into a different data format
to ensure readability~\cite{Lekkas2004,Schmidt2005}. These problems, though outside of the scope of our concept
proper, should be mitigated by a modular design and use of openly documented
technology.
\Section{System design and implementation}\label{sec:design-impl}
In this section, we describe  a high-level architecture for a voice archive scenario,
the basic concepts underlying our approach,
and a concrete implementation of its central component which ensures 
security of the archive.
\SubSection{Architecture design}\label{sec:architecture-design}
The main design principle in the implementation and deployment variant we describe here is
that of minimal technical requirements at the part of the communication clients.
The presented architecture is also used as a system model for potential attacks in the next section.
\begin{figure}[ht]
\centering
\ifpdf
 \resizebox{0.45\textwidth}{!}{\includegraphics{sysmodel.pdf}}
\else
  \resizebox{0.45\textwidth}{!}{\includegraphics{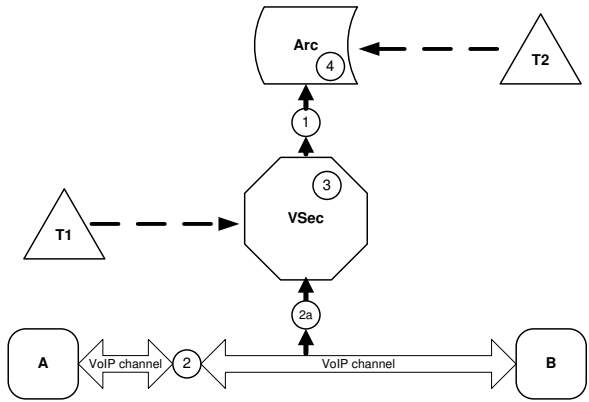}}
\fi
\caption{High-level architecture\label{fig:sysmodel}}
\end{figure}
Figure~\ref{fig:sysmodel} shows the communication between two partners \CA and \CB over
a VoIP channel, which is an idealisation, comprising in particular session initiation and communication. 
At a certain point in the channel, the VoIP security component \VS listens
to the communication. \VS, which is the main component implementing the base 
security concept of Section~\ref{sec:base-concept}, could be located at the site of either
of the parties \CA or \CB, e.g., in the case of call-centre applications, but this
is not necessary in principle.
The component \Arc denotes the archive to which the secured VoIP communication is submitted and
then persistently stored. 
\TTA and \TTB are additional time-stamping authorities (TSA) which come into play to raise
resilience against attacks exerted by attackers situated in positions
(1)-(4), see Section~\ref{sec:secur-cons} below.

\Arc is the component to which also the mentioned long-term aspects are deferred.
Research has yielded some sound solutions for the efficient, secure, long-term
archiving of signed digital documents, mostly based on the use of
hash-trees~\cite{Merkle1990}. Some systems are already on the market which can be 
flexibly combined with~\cite{Viebeg2006,ArchiRefPage} or are already integrated into document
management systems~\cite{OpenText}. Application of such systems in the role
of the voice archivist \Arc suggests itself.

\VS will often be under the control of one of the parties or even be integrated in their
VoIP infrastructure. Neither the exact position in the channel nor the technical
method by which \VS intercepts it is essential for the architecture and its security
properties discussed below in Section~\ref{sec:secur-cons}. 
The role of \VS can be passive or dual, listening to communication and enforcing
policies on it. We will present a single example of policy enforcement by
\VS in Section~\ref{sec:base-concept}. The separation of such a component
from the rest of the system is standard in security engineering where it is
commonly known as a reference monitor~\cite{refmon}.

In accordance with the mentioned principle of minimal client involvement the 
communication between \CA and \CB is in particular not required to be digital, let alone SIP/RTP based,
end-to-end, provided that there is some part of the channel which is VoIP. This condition is 
already met in many mobile and public switched networks. 
In particular, the phones used by \CA and \CB need not be ISDN or VoIP phones.

Furthermore, the application domain of the presented voice archive concept and architecture
is not restricted to  intra- and inter-organisational telephony.
In principle it comprises archiving of \emph{any packet-based digital voice conversation}, e.g., 
in (tele-) conferences, or digital radio communication as used by government authorities
and organisations entrusted with security tasks, see~\cite{Klappenbach2004}.%
\SubSection{Base concept}\label{sec:base-concept}
A digital multimedia communication consists in general in two channels 
transporting data packets and meta data back and forth between
two partners. 
The proposed concept handles these data by creating so-called intervals,
containing several packets. 
The  integrity of each interval is secured by hashes and the application
of a cryptographic secret to protect the hash value.
At the start of an archiving process \VS cryptographically secures
sufficient data to provide a unique identity the subsequent stream, to ensure
the archive's integrity.

To ensure cohesion two measures are taken. First
a cryptographic chain is formed by including the hash 
value of one interval in the data used to compute its successor. An attacker
cannot remove a single interval without invalidating the
subsequent hashes and thus being detected. 
Due to the cryptographic secret applied to each interval 
any manipulation of a interval's  content as well as addition or
deletion of intervals is excluded.

At the end of the chain a special terminating package is 
added signalling its end. If this package is missing it is
clear that the archived version was tampered with in transmission
between \VS and \Arc or was incompletely archived due to a malfunction
in \Arc.

During archiving of a conversation, \VS monitors the quality of service
(QoS) of the voice connection. 
If a certain QoS threshold in terms of, e.g., packet loss, 
is under-run then either the connection
quality is poor and the participants cannot understand each other
with a sufficient quality, or there is an ongoing attempt to
attack the communication. 
Both cases lead to a lack of
trustworthiness of the archived conversation.
It is now a matter of policy how to deal with this QoS under-run. 
It could be ignored, the users could be notified while continuing the archiving, 
the archiving could be aborted, or the call could be terminated. 
The first two options open the path for certain semantic attacks as will be discussed in
Section~\ref{sec:secur-cons}. 
We favour termination of the call as this the option for maximum security 
and the QoS threshold is seldom reached without a breakdown of the connection 
anyway due to insufficient understandability or software timeouts.
It should be noted that this is not an essential design decision and that 
the policy actually employed depends on security requirements of the 
application scenario. QoS as such and in particular packet loss is, 
however, an essential point that a secure VoIP archive has to cope with.

The QoS observed by \VS is not identical to the ones perceived by either \CA or \CB,
but rather only an upper bound for them. The sharpness of the bound depends on the
``distance'' of \VS to \CA or \CB in the channel. In applications it can therefore 
be a good choice to integrate \VS with the VoIP system of one of the parties, perhaps
the one interested in the archiving in the first place.

The length of a interval and the QoS threshold are the main free parameters in the concept.
While the latter depends on application-dependent (security) requirements, see~\cite{Athina2002}, 
the former should be optimised so as to minimise cryptographic workload and storage overhead.
This must be balanced with the loss of conversation context in the case of a QoS under-run, 
when the last interval has to be discarded. Dynamic adaption of interval length is also the
lever to satisfy the scalability requirement described in Section~\ref{sec:requ-secure-arch}.
\SubSection{Implementation and protocol integration} \label{sec:impl}
\subsubsection{Overview}
The archive system has been implemented as a prototype and tested with several
soft phones and hardware- devices (e.g., AVM's Fritz!Box~\cite{Fritz}) using the SIP and RTP protocols.
For nomenclature of these protocols used below see~\cite{rfc2327,rfc2543,rfc3261}.
In the place of \CB we used mobile phones, ISDN phones, and also SIP software clients.
\VS was implemented using C\#, running on an embedded x86 based PC without keyboard,
mouse or video ports. The Linux operating system with the Mono-framework was used to run
the program. It was placed as a proxy between \CA and the Internet using its two 
NICs, and thus supports multiple clients and calls at the same time.
\VS is implemented as an outbound proxy substituting A's original outbound proxy.
The proxy modifies RTP ports and IP addresses contained in the SIP packets redirecting 
them to itself and in turn forwards them to the original recipients.
A traditional PC was used for \Arc, connected using the third NIC of \VS. They communicate over
a reliable TCP channel (for privacy also the TLS protocol could be used).

RTP-packets are grouped in intervals where each interval is signed and stored on \Arc. Each 
interval consists of about one second worth of RTP packets. Because the implementation 
only supports bidirectional calls (conference calls have not yet been
implemented at the time of writing), there were two intervals per second, one for each 
channel/RTP stream. The duration of an interval is one of the main configuration
parameters to be tuned. One second proved to be sufficient to provide a high 
level of security for the context of the talk on the one hand.
On the other hand it keeps the computational power required
by far low enough for the used x86 processor and also the storage overhead 
(400 bytes for PKCS\#7 signatures without embedded certificates)
to payload ratio small. 

\VS carries a X.509 certificate together with the 
private (RSA) key to sign (using asymmetric cryptography)
all intervals, including the special start and stop interval 
containing meta data.
The certificate that \VS carries is not only used to sign the contents of the
call for the final archiving and later verification, but also to authenticate \VS to \Arc. 
Immediately after completion intervals are transmitted to \Arc, which 
then performs several tests on the interval, including verification of the signature and then stores 
it as chunks into an open file.
The executed tests are:
\begin{enumerate}
  \item[CHK1] Checking whether the first interval with the meta data is correctly signed by the
        external time-stamping service \TTA. In particular \Arc compares the  time with that recorded
        by \VS in the interval preventing \VS (without collaboration of \TTA) from forge a different call time.
        If there is an additional third party proof of the time of communication (like
        an itemised bill from the phone company) this can as well be compared to this 
        initial time stamp.
  \item[CHK2] Validating the PKCS\#7 signature of the interval. This authenticates
        \VS to \Arc and ensures that no other person can submit streams to \Arc.
        \Arc does not know the private key that \VS knows, but can check it against the
        certificate and a trusted root.
  \item[CHK3] Verifying interval chaining. \Arc stores the
        SHA1 hash of the last interval and compares it to the embedded hash value in the
        current one. If they do not match the chain was broken and communication is terminated.
  \item[CHK4] Checking packet loss by checking the absolute sequence numbers in the interval structure.
        If the packet loss is above the QoS threshold, the archiving is aborted
        by \Arc and the call terminated by \VS, by injecting a BYE command terminating SIP and RTP forwarding.
        We chose 1\% packet loss as threshold, 
        which still ensures good understandability.
  \item[CHK5] Checking the time embedded by \VS in the intervals whether it drifted not more than
        two times the interval duration from the internal clock of \Arc. In the demonstrator,
        clocks were synchronised with NTP, which should be replaced with a secure, trustworthy time source
        in a production system. 
  \item[CHK6] Checking the temporal integrity of the RTP packets, i.e., whether the time-stamps and 
        sequence numbers stored in the RTP protocol, 
        which can suffer from overflows and rollovers, are consistent with the time recorded in 
        the interval.
\end{enumerate}
In this way the whole conversation is continuously and securely streamed from \VS to \Arc and 
\VS never needs to store more than 2s worth of RTP packets per call
in memory (in particular a hard disk is in principle not required on \VS). 
\VS also has to do only about two RSA signing operation per second.
\subsubsection{Data format}
The format used to stream the call from \VS to \Arc consists of intervals, including
a special initial interval carrying meta data, a final interval containing the reason
of the termination of the call, and several intervals with voice data (initial and final interval
do not carry any voice data). 
Each of the intervals stems from either RTP channel from \CA to \CB or the other
direction. Each interval is embedded in an PKCS\#7 signed envelope container. 
Only the first interval's PKCS\#7 signed envelope container contains the whole
certificate chain up to, but not including the root, while all other containers
don't need to carry this redundant information.
The first interval is also additionally wrapped in a signature
from the time stamp service \TTA.
\Arc simply stores each interval (together with its signature), e.g., to its hard drive 
after performing the described checks on it. 
\begin{figure}[t!]
  \centering
\ifpdf
  \resizebox{0.475\textwidth}{!}{\includegraphics{interval_chain.pdf}}
\else
  \resizebox{0.475\textwidth}{!}{\includegraphics{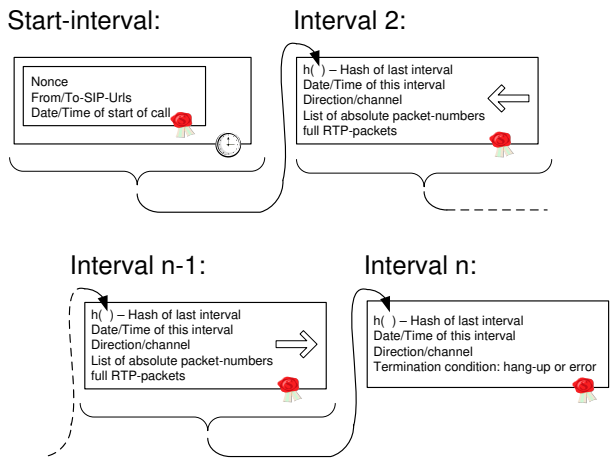}}
\fi
  \caption{Format of the archived calls showing chaining of interleaved channels\label{fig:interval-chain}}
\end{figure}
The signed and timestamped content of the initial interval consist of 
\begin{itemize}
	\item A random nonce. This mitigates a replay and duplication attack, see Section~\ref{sec:secur-cons}. 
	\item The date and time of the call.
	\item The from and to SIP URLs of the caller and callee.
	\item The mapping of RTP payloads to the used media formats and codecs. This information
	      is embedded in the SDP bodies of, e.g., the INVITE request from the SIP signalling.
	      Without this information the dynamic RTP payload in the rage from 96--127 would
	      not be known later when the archived call is played back. As an implementation
	      variant it is also possible that the implementation modifies the SDP negotiation
	      to only allow certain codecs which have a high probability to have existing 
	      implementations available later (e.g., no proprietary codecs).
\end{itemize}
The signed content of the last interval consists of
\begin{itemize}
	\item Hash of the second-to-last interval to complete the cryptographic chain.
	\item A flag that this is the last interval of an archived call.
	\item Reason for termination: Protocol or network error, 
              regular hang up by \CA or \CB, violation of packet 
              loss QoS threshold under-run, or tamper detection.
\end{itemize}
All other intervals contain actual speech data, as follows.
\begin{itemize}
	\item Hash of the complete, signed (and time-stamped in the case of interval no.~2) 
              interval before this interval.
	\item Date and time of this interval.
	\item Direction/channel of this interval. In the implementation for duplex calls between
	      only two parties this can be from \CA to \CB or the other direction.
	\item List of absolute sequence numbers of the contained RTP packets.
	\item The complete RTP packets referenced by this list, including their payload type and
	      the truncated timestamps and sequence-numbers.
\end{itemize}
All intervals together form a cryptographic chain from the first to the last interval. 
Intervals of both directions (channels of the duplex phone call) are interleaved. 
After two intervals of one direction
there \emph{must} be a interval with the other direction and the date and time in the intervals must
be sequential. Otherwise \Arc or any other verifier of an archived conversation has
to reject the file, just as if a hash value or signature was invalid.
\subsubsection{Replay window}
\VS contains a component to sort incoming packets and has a replay protection which removes replayed
packets. This is implemented using a 32 Bit sliding window. Because RTP-packets only contain
a 16 Bit-sequence number (which is even recommended to not start with 0 or 1, but with a random
value to improve symmetric encryption) and only a 32 Bit time stamp (which also starts at
a random value, can overflow, and is not related to an absolute time) this component
also helps in creating absolute sequence-numbers starting at 0 and checking the system time
against the time stamp of the packet for consistency. If any of these checks is violated, the
call is terminated. If \VS does not detect these things (because it has been tampered with), 
then \Arc will stop archiving because it employs the same tests.
\Section{Security considerations} \label{sec:secur-cons}
We discuss some general issues before we come to particular attacks.
First, the security of the digital voice archive cannot be better than that of an
analog archive. Forgery of voice communication is generally considered 
difficult, however a cunning attacker with sufficient
resources might eventually be able to simulate any VoIP stream he desires, and 
insert it to the archive, e.g., from the position (2a) in Figure~\ref{fig:sysmodel}.
Such attacks are out of the scope of the present concept, and will remain so without 
involvement of the clients and in particular their authentication and provable trustworthiness.
Nevertheless, the base principles and architectural design of the voice archive
can prevent the insertion of forged communications through other channels, respectively,
from other positionings in the system model, as will be shown below.

Finally, an inherent feature of the base concept described in Section~\ref{sec:base-concept} 
is its fragility. Artificially lowering the quality of the connection, e.g., by causing
or simulating (e.g., by B) packet loss, is clearly an attack vector and poses a risk to
the trustworthiness of a  voice archive.
When the voice quality in a interval falls below a given threshold, the cryptographic chaining
ends, providing a Sollbruchstelle (predetermined break point) for the probative value of the
archived communication, since only the parts before that break are cryptographically verifiable.
In contrast, most other schemes for securing the integrity of streamed data, e.g., the
signing method of~\cite{Perrig2000} aim at loss-tolerance, for instance 
allowing for the verification of the stream signature with some probability, even in
the presence of intermittent packet loss. We argue that for the probative value of
inter-personal, natural language communication, the former behaviour is advantageous.
An archived call with an intermediate one-second gap can always give rise to speculations
over alternatives of filling the gap, which are restricted by syntax and grammar, but
can lead to different semantics. Using this, a clever and manipulative attacker could
delete parts of the communication before they enter the archive, claiming with some credibility
that the remnants have another meaning than intended by the communication partner(s).
But if the contents of a conversation after such an intentional deletion are unverifiable and
thus cannot be used to prove anything, this kind of attack is effectively impeded.

We now perform an attack analysis by the location of a potential attacker within the system
described in the previous section. The protection targets under attack are mentioned 
for each single one. The numbering of the attacks corresponds to the positions marked in 
Figure~\ref{fig:sysmodel}. The analysis rests on the abstract concept of Section~\ref{sec:base-concept}
to exhibit its salient security-related features. The technical variants chosen
in Section~\ref{sec:impl} are mentioned in the end of this section.

\noindent\textbf{(1) Man-in-the-middle} (between \VS and \Arc).
The main threat against our method to securely archive voice data emanates from an attacker
who can intercept and manipulate the data in transmission from \VS to \Arc.
This attack vector makes the strong assumption that any transport layer security
between \VS and \Arc has been broken.
There are two ways to insert a forged conversation into \Arc; either he interrupts an
ongoing submission to \Arc by suppressing \VS after, say, interval $n$ and continues it himself
with interval $n+1$; or, he starts a submission to \Arc himself, pretending to be \VS.
These options, 
threatening the cohesion of conversations, respectively, the integrity of the archive,
 are ruled out by the base concept underlying \VS, depending on two particular
countermeasures.
\begin{enumerate}
\item[C1.1] Build a chain of cryptographically secured hash values using a secret known
only to \VS (or shared between \VS and \Arc). This prevents (1) from executing the first
attack variant since he cannot continue the chain of intervals. For detectability of
(1) by \Arc during a submission, the use of the correct secret should be verifiable by \Arc.
Various symmetric or asymmetric cryptographic methods provide for the desired features. 
\item[C1.2] To start the chain of secured intervals and prevent the second attack variant by (1), 
some initial data must be secured.
In the simplest case, this is not different from C1.1 and consists only in securing the first 
interval of the transmission. We will see below that refinements are desirable to
heighten resilience of the overall system.
The initialisation can be used in close conjunction with C1.1, e.g., if the Diffie-Hellman
protocol~\cite{Diffie1976} is used to establish a shared secret. On the other extreme part of the 
spectrum, \VS can use certificate-based authentication toward \Arc to initiate the submission.
\end{enumerate}
It should be noted that these two countermeasures are inherent in our original concept.
All other countermeasures below depend on them as technical or process implementation
variants providing gradually improvements on security.

\noindent\textbf{(2) Replay forgery.}
Consider an attacker listening to the VoIP channel at some point near \CA or \CB and able
to simulate a VoIP conversation toward \VS. 
Assume (2) is recording some initial part of a real archived conversation between \CA and \CB.
This attacker can then in principle compromise the integrity of \Arc by
replaying this initial piece to \VS and continuing it with a different
conversation, either forged synthetically, or in collaboration with \CA and \CB.
In consequence, non-repudiation of the fact that a particular conversation has
taken place is easier, since now \Arc contains two data sets with very similar, or even
bitwise identical initial sections of a certain length.  
The threat of this attack to the archive's integrity is even larger in the following, 
related case

\noindent\textbf{(2a) Replay forgery with VoIP source control.}
Due to the imperfection of VoIP communication, the attack of (2) is generally difficult since 
he does not know precisely which packets actually arrived at \VS. The tolerable packet loss
of VoIP is therefore likely to entail discrepancies between original and replay 
even in the first interval of a communication, depending on interval size.
Therefore position (2a), where he directly listens to \VS's data source
is better suited to a replay attack, in particular
since the two conversations will even
appear to have been carried out at the same time and along the same route since (2) 
replays the recorded SIP meta-data as well. This meta-data, containing routing information
and system times is also difficult to forge in position (2).
\begin{enumerate}
\item[C2.1] Exploitation of randomness of VoIP, i.e., packet loss is the simplest countermeasure
against (2). If the interval size is large enough the mentioned discrepancies are likely to
occur, in effect discriminating original from replay forgery. To raise this likelihood,
\VS could secure a determined number $\gg 1$ of intervals to initiate the chain.
\item[C2.2] Using a random seed to initiate the secured chain is a much stronger 
countermeasure. Again, various methods can be applied to that end, either within
\VS or using an external source of trust, see C3.2 below.
\end{enumerate}
Note that the question whether a data set that is detected to be an artifact of a
replay attack --- something which can always be determined only with a certain 
probability --- should be (marked and) kept in \Arc or discarded is a matter of policies.

\noindent\textbf{(3) Compromised \VS secret.}
If an attacker has control over \VS to the extent that she knows the secret used to initiate
and build a chain of intervals she is in the position, like (1) but more effectively, to insert
forged conversations into \Arc. If we assume that (3), possessing, e.g., \VS's authentication data,
completely appears as \VS to \Arc, still two categories of countermeasures apply.
\begin{enumerate}
\item[C3.1] Internally, \VS can use rotation of the secrets, e.g., one-time keys generated
for every conversation from a master secret, which in turn requires a significantly better
protection. The hardware security provided by trusted platforms suggests itself for this purpose.
\item[C3.2] An external source of trust can be invoked at every submission to initiate the
chain of intervals. For instance, a time-stamping service \TTA can be used to sign the data
initiating the chain. This relies of course on the assumption that the authentication
of \VS with respect to \TTA remains unbroken and thus (3) cannot obtain time stamps on her own.
\end{enumerate}
C3.2 additionally raises resilience against replay attacks. Regarding (3), her only
remaining attack if the latter countermeasure is in effect is a replay attack as 
well, which will easily be detected due to identical time stamps over, and random seeds in,
the initial data.

\noindent\textbf{(4) Forgery by the archivist} is prevented by a design principle
of the presented system.
\begin{enumerate}
\item[C4.1] The separation of duties between \VS and \Arc makes it difficult for \Arc
to forge secured conversations and to claim that they originate from \VS. This holds always
in the asymmetric situation that \Arc does not know the secret \VS uses to secure chains 
of intervals, but is able to verify them, which remains a prerequisite of the secure archiving.
Even if (4) has the additional power of (3) the further separation of duties by C3.2 and the
invocation of the time stamping service \TTA effectively suppresses his ability to 
forge conversations that ``look like'' those coming from \VS.
\item[C4.2] One basic security problem of long-term archiving remains, namely the authority
of \Arc over the archived data. This enables \Arc at least in principle, given enough time
and computational power, to manipulate the archived data.
A method that is considered to be generally effective for mitigating this threat is the
use of periodic time-stamps from another time-stamping authority \TTB 
over the bulk of data archived during a period.
Hash-trees are the method of choice to implement this process effectively.
\end{enumerate}

Special attacks can be attempted in a combination of the positionings above.
For instance, an attacker combining roles (1) and (2a) could try to insert a back-dated
call into \Arc. From (2a), he would insert calls at times of his choosing into \VS, intercept and 
store the generated initial intervals containing time-stamps from \TTA at (1) and suppress transmission of 
the conversation to \Arc. Later he would try to insert the call he wishes to back-date from
(2a) into \VS, but replacing, from (1), only the initial interval by a stored one time-stamped at the
desired earlier time. This fails due to the chaining of intervals and the uniqueness
of initial intervals induced by the use of a nonce, i.e., this attack is suppressed by 
C2.2, in conjunction with C1.1 and C1.2.

In the technical variant described in Section~\ref{sec:impl}, C1.2 is 
implemented by using digital signatures based on asymmetric cryptography on all intervals
and enveloping the first interval in a cryptographic time-stamp.
Countermeasures C2.2, and C3.2 are taken into account by including a random nonce
in the first interval, and use of the initial time-stamp from the external TSA T1, respectively.
\Section{Conclusions}\label{sec:conc}
We have presented a system for archiving VoIP-based communication which has some salient
security features in contrast to existing digital voice recording solutions.
The present solution is stand-alone and offers a high degree of scalability, 
ease of integration,  and efficiency without trade-offs with respect to security.

Certain advanced implementation variants can be envisaged based on our concept.
In particular, utilisation of trusted platforms (TP) as specified by the trusted computing group~\cite{TCG12}
suggest themselves.
A TP can be used for various security-related tasks in \VS, e.g., storing
secrets, securing data channels and interfaces, or providing a trustworthy computing environment.
As another instance of trusted computing usage, the time-stamping by \VS could be implemented using an
internal trusted clock of \VS seeded daily by \TTA, in order to reduce the cost of purchasing 
time-stamps.

A real-world implementation also needs to consider conditions for, and signalling and negotiation of
recording of a conversation. The draft standard~\cite{SIP-Pre} describes a method by which
``One party may assert either their desire to record or their restriction of the other party's recording''.
Using these assertions in our archiving architecture in the sense that \VS evaluates
and respects them would be a nice way to disarm privacy reservations to indiscriminate recording
of calls. On the other hand, signalling of archiving status and (reasons for) termination of the 
archiving, respectively, the call are desirable future features. A device independent
way using speech synthesis can be envisaged.

As an outlook, it seems possible to extend the present concept to a full-fledged
electronic signature over VoIP-based conversations.
This includes, either unilateral or mutual, authentication of communication partners, 
non-repudiation of a conversation's content, and ultimately leads to a probative force
of such conversations equivalent to other electronically signed documents.
This enables declarations of will and establishment of binding contracts by voice.
Of course such an advanced scenario is no longer possible without a certain involvement
of the clients, in particular their trustworthiness and integration into an authentication
infrastructure like a PKI. Working out this advanced scenario is in progress~\cite{PATENT_PENDING}.
\end{document}